\documentclass[prd,twocolumn,nofootinbib]{revtex4-2}
\usepackage{graphicx} % Required for inserting images
\usepackage{amsmath}
 \usepackage{natbib}
\usepackage{threeparttable}
\usepackage{pgfplots}
\usepackage{multirow}
\usepackage{array}
\pgfplotsset{compat=1.18}
\usepackage{caption}
\usepackage{subcaption}
\usepackage{adjustbox}
\usepackage{tabularx}
\usepackage{hyperref}
\usepackage{float}
\usepackage{newunicodechar}
\newunicodechar{−}{\ensuremath{-}}
\usepackage{booktabs}
\usepackage{slashed,epsfig,amsmath,amssymb,enumitem} \usepackage[papersize={8.5in,11in}]{geometry}
   \usepackage{bm}
\usepackage{graphicx}
\newcommand{\be}{\begin{equation}}
\newcommand{\ee}{\end{equation}}

\usepackage[utf8]{inputenc}
\begin{document}

\title{Comment on a ``Comment on ``Standard Model Mass Spectrum and Interactions In The Holomorphic Unified Field Theory""}

\author{J.~W.~Moffat\textsuperscript{1} E.~J.~Thompson\textsuperscript{1,2}}
\affiliation{[1]Perimeter Institute for Theoretical Physics, Waterloo, Ontario N2L 2Y5, Canada}
\affiliation{[2]Department of Physics and Astronomy, Trent University, Peterborough, Ontario K9L 0G2, Canada}

\maketitle

\small

The comment \cite{Cline:Comment} claims that our nonlocal form factor induces a one-loop photon mass because the regulator function is not gauge invariant. This conclusion follows from an incomplete nonlocalization omitting the required contact vertices and the functional-measure term and therefore violates the nonlocal Ward identity.

\maketitle

\section{What the Comment computes}
Equation (1) of the Comment sets external momentum to zero and inserts the nonlocal factor directly in the fermion loop:
\begin{equation}
\Pi_{\mu\nu}(0)
= e^2\!\int\!\frac{d^4p}{(2\pi)^4}\,
\frac{\mathrm{tr}\big[\gamma_\mu(\slashed{p}+m)\gamma_\nu(\slashed{p}+m)\big]}
{(p^2-m^2+i\epsilon)^2}\,
e^{+2p^2/M_*^2}
\end{equation}
\begin{equation*}
\;\sim\; \frac{\alpha}{4\pi}\,M_*^2\,\eta_{\mu\nu},
\end{equation*}
and concludes that a photon mass is generated. This calculation neglects the nonlocal gauge symmetry, the associated higher-point contact vertices, and the compensating measure term that arise in the consistent, gauge-invariant nonlocalization. 

\section{Gauge-invariant nonlocal QED in a nutshell}
A consistent nonlocalization of QED proceeds in two steps, at the classical level, first decorate only the interaction part with entire smearing operators, and then add specific higher-point contact terms so that tree amplitudes obey decoupling and an on-shell nonlocal $U(1)$ symmetry. The fermion transformation is a field-dependent, nonlocal operator or a quantum representation, while $\delta A_\mu=-\partial_\mu\theta$ remains standard. This restores tree-level Ward identities and Compton decoupling order-by-order. 

At the quantum level, in the functional formalism, the fermion measures are not invariant under the deformed symmetry. The remedy is an explicitly computable measure action $S_{\rm meas}[A]$. This term is required to maintain the Ward identity in loops and it preserves finiteness and Poincaré invariance. 

At one loop there are three contributions to  the vacuum polarization from the nonlocally deformed classical vertices and one from the measure action. The first two contain longitudinal pieces, the measure diagram cancels these exactly, restoring transversality:
\begin{align}
p_\mu\,\Pi^{\mu\nu}(p)=0,\\\qquad
\Pi^{\mu\nu}(p)=\big(p^2\eta^{\mu\nu}-p^\mu p^\nu\big)\,\Pi_T(p^2).
\end{align}
This is shown explicitly by adding the three graphs see Eqs.\ (3.31)--(3.36) and Fig.\ 4 in \cite{Moffat1991}. The total carries an overall factor of $p^2$, so that
$\Pi^{\mu\nu}(p)\to 0$ as $p^2\to 0$ showing no photon mass is generated. In closed form, the transverse scalar can be written as Schwinger parameters in terms of an $E_1$ exponential integral, the small $p^2$ behaviour is logarithmic and matches the usual dimensional regularization result, with no quadratic mass term.

For the photon vacuum polarization a transverse $\Pi(p^2)$ gives:

\begin{align*}
\Pi^{T}(p^2) &= -\,\frac{e^2 p^2}{2\pi^2} \,
\exp\!\left( -\frac{p^2}{\Lambda^2} \right)
\end{align*}
\begin{equation}
  \times  \int_{0}^{1/2} \! dx \; x(1-x) \,
E_1\!\left[\, x \frac{p^2}{\Lambda^2}+\frac{1}{1-x} \frac{m^2}{\Lambda^2} \,\right] .
\end{equation}
The general structure at one loop for any 1PI amplitude $\Gamma(1)$, the nonlocal regulator yields the same logarithmic UV behavior as dimensional regularization, given by Eq.~3.36:
\begin{equation}
   \Pi^{T}(p^2) 
= -\frac{e^2 p^2}{2\pi^2}[ \frac{1}{6} \frac{2}{4-D} - \frac{1}{6}\gamma - \frac{1}{6} \text{ln}(2\pi)
\end{equation}
\begin{equation}
    - \int_{0}^{1} \! dx \; x(1-x) \ln\!\big( x(1-x)p^2 + m^2 \big) + \mathcal{O}(4-D)].
\end{equation}
This follows from the correspondence:
\begin{equation}
    \frac{2}{4-D}\sim \text{ln}(\Lambda^2).
\end{equation}
The choice of coefficients of $\Pi^T(p^2)$ of the logarithm depend on the renormalization scheme. Moreover, the $\beta$-functions and the nonlocal dimensional regulation are the same. Power divergences that would be gauge-violating in a naive cutoff do not appear here; the entire-function and exact gauge invariance forbid a photon mass term and maintain Ward identities.

\section{Sign of the exponential}
The Comment’s Eq.~(1) inserts the nonlocal factor $\exp(+\,2p^2/M_*^2)$ directly into the Minkowski-space fermion loop, while also stating that Wick-rotated loop diagrams are used.  This is inconsistent as after Wick rotation, the exponent must carry a negative sign to produce the intended ultraviolet suppression.

In our framework, the nonlocal form factor in position space is:
\begin{equation}
F\!\left(\frac{\Box}{M_*^2}\right) = \exp\!\left(-\frac{\Box}{M_*^2}\right).
\label{eq:formfactor}
\end{equation}
With Minkowski signature $(+,-,-,-)$, the d’Alembertian acts on a plane wave $e^{-ip\cdot x}$ as:
\begin{equation}
\Box \, e^{-ip\cdot x} = -\,p^2 \, e^{-ip\cdot x}, \qquad p^2 = p_0^2 - \mathbf{p}^2 .
\end{equation}
Thus in Minkowski momentum space, Eq.~\eqref{eq:formfactor} becomes:
\begin{equation}
e^{-\Box/M_*^2} \;\longrightarrow\; e^{+\,p^2/M_*^2}.
\label{eq:minkexp}
\end{equation}
The quantity $p^2$ here is not positive definite, for timelike momenta $(p_0^2>\mathbf{p}^2)$ it is positive, for spacelike momenta $(p_0^2<\mathbf{p}^2)$ it is negative.  Consequently, Eq.~\eqref{eq:minkexp} does not give uniform suppression in Minkowski space and cannot be used directly for UV finiteness proofs.

The standard procedure, followed is to Wick rotate the contour $p^0 \to i p_E^0$, which sends:
\begin{equation}
p^2 \;\longrightarrow\; -\,p_E^2, \qquad p_E^2 \equiv p_E^{0\,2} + \mathbf{p}^2 \ge 0 .
\end{equation}
Here $p_E^2$ is positive definite, and Eq.~\eqref{eq:minkexp} becomes:
\begin{equation}
e^{+\,p^2/M_*^2} \;\xrightarrow{\text{Wick}}\; e^{-\,p_E^2/M_*^2} ,
\end{equation}
which gives exponential damping for all Euclidean momenta.  This is the origin of UV finiteness: in Euclidean space the large-$p_E$ region is uniformly suppressed.

The Comment’s Eq.~(1) keeps the Minkowski sign $e^{+\,2p^2/M_*^2}$ while also invoking a Wick-rotated evaluation.  This is a sign error, once the rotation is made, the factor is $e^{-\,2p_E^2/M_*^2}$, so there is no exponential growth, and the integral is well-defined.  Moreover, when the required contact vertices and the fermionic measure contribution are included, the full $\Pi_{\mu\nu}(p)$ is transverse and satisfies $\Pi_{\mu\nu}(0)=0$, leaving the photon massless.

\section{Where the Comment’s mass term comes from}
Dropping the contact vertices and $S_{\rm meas}$ breaks the nonlocal Ward identity and allows a spurious, momentum-independent $\eta_{\mu\nu}$ term at $p=0$. The 1991 construction shows these missing pieces are not optional. They are uniquely fixed up to gauge-invariant functionals of $F_{\mu\nu}$ by demanding decoupling and gauge invariance of the functional integral. With them included, the longitudinal parts cancel and no photon mass arises. 

\section{Implementation in HUFT}
Our HUFT papers employ the same entire-function form factors damping $e^{-p_E^2/M_*^2}$ in Euclidean space but, crucially, the gauge sector is nonlocalized by the gauge-invariant procedure summarized above as we add the required contact interactions to restore tree-level decoupling and on-shell symmetry, and include the fermionic measure term that enforces the nonlocal Ward identities at loop level. The photon remains massless to one loop and order-by-order, and $\Pi^{\mu\nu}$ is transverse \cite{MT2025}. The Comment’s estimate corresponds to using only naively smeared vertices and therefore does not apply to HUFT’s gauge sector.

\end{document}